\UseRawInputEncoding
\documentclass[aps,prl,reprint]{revtex4-1} 

\usepackage{amsmath}    
\usepackage{graphics}   
\usepackage{color}      
\usepackage{subfigure,epsf}  
\newcommand{\beq}{\begin{equation}}
\newcommand{\ee}{\end{equation}}
\newcommand{\beqa}{\begin{eqnarray}}
\newcommand{\eea}{\end{eqnarray}}
\newcommand{\ie}{{\textit i.e. \ }}
\newcommand{\eg}{{\textit e.g. \ }}

\begin{document}  



\title{
Tension-clock control of human mitotic chromosome oscillations
}

\author{Nigel J. Burroughs$*$}
  \affiliation{Mathematics Institute, University of Warwick, CV4 7AL. UK.\\
  $*$ Corresponding author: N.J.Burroughs$@$warwick.ac.uk}
\author{Andrew McAinsh}
\affiliation{Warwick Medical School, University of Warwick, CV4 7AL. UK.}

\maketitle


{\bf 
  During cell division paired (sister) chromosomes are
  observed to perform approximate saw-tooth oscillations across the
  cell mid-plane.  Experimental data suggests that these
  oscillations are regulated through intersister tension.
  We propose a time dependent tension threshold model
   that exhibits three stable periodic solutions and the phase diagram can be generated semi-analytically. 
 Incorporation of diffusive noise 
  reproduces realistic oscillations,  with realistic periods, amplitudes and reproducing
   the observation that 
 either sister of the pair can switch direction first.  
  }

\vskip 0.5cm

High fidelity cell division in mammals relies on retaining duplicated chromosomes (called chromatids) together in a holding pattern near the cell equator until all paired chromatids are correctly attached to the spindle. In particular,  sisters needs to be attached by microtubules to  separate spindle poles \cite{mcintosh2012,nenad2016}. 
Whilst in this holding pattern, chromatid pairs are observed to oscillate parallel to the spindle axis in the  majority of animal cells, including humans \cite{Skibbens1993,Jaqaman2010,Wan2012mbc}. 
Oscillations are pseudo-periodic with sections of near constant speed separated by a directional switching event whereby both of the sisters  switch direction, Fig. 1A.
Oscillations are predictive of segregation errors, \cite{Sen2021} and are attenuated in cancer cells  { \cite{lemura2021}.}

A number of mathematical models of chromosome oscillations have been proposed incorporating various mechanisms, reviewed in \cite{Vladimirou2011,Civelekoglu2014},
oscillations typically arising from mechanical processes \cite{Joglekar2002,Civelekoglu2013,Banigan2015,Klemm2018,Schwietert2020} or chemical feedbacks \cite{liu2007,liu2008,shtylla2010,cabrerafernandez2018,Alejandro2021}. 
Microtubules are intrinsically dynamic, exhibiting dynamic instability \cite{Mitchison1984} which  provides a natural driver for oscillations. 
Collective modelling of  microtubule binding with force dependent catastrophe/rescue effectively synchronises  dynamic instability of the microtubule bundle \cite{Joglekar2002,Banigan2015,Schwietert2020}.





However, these tug-of-war models are 
inconsistent with the low intersister force inferred from data \cite{Armond2015plos}, suggesting 
that directional switching is  controlled. 
The regulatory processes underpinning this control and generates  oscillations remain poorly understood, but believed to be a consequence of sister-sister coregulation mediated by  tension in the spring-like centromeric chromatin that connects the sisters  \cite{RiederSalmon94,Burroughs2015elife,Wan2012mbc,rago2013}.
Sub-pixel 3+1D imaging of kinetochore positions has shown that the two sisters rarely switch coincidently, with both switching orders  being observed, \cite{Burroughs2015elife}. Specifically  the leading sister  or the trailing sister can switch first, referred to as
leading/trailing sister initiated directional switching, LIDS, TIDS, respectively, \cite{Burroughs2015elife}.
This results in a relaxation (LIDS) or  overstretch (TIDS) respectively of the centromeric chromatin spring at a directional switch;
i.e. there is a phase where both sisters are attached to polymerising (LIDS), respectively depolymerising (TIDS) microtubules, Fig. 1B, \cite{Armond2015plos}. This raises the question of how control of chromatid direction through the spring tension  can give rise to oscillations where both of these states can occur with seemingly little impact on the global oscillation  characteristics, including the oscillation period and amplitude. 
We propose a new mathematical model, formulating the tension clock hypothesis of \cite{Burroughs2015elife}. It  is a force dependent control model, where a dynamic tension threshold controls the time of directional switching of the attached microtubules.



We model the movement of 
the kinetochores, multi-protein machines that assemble near the middle of  each sister chromatid. Kinetochores  facilitate the mechanical attachment of the chromosome to a bundle of microtubules, called the K-fiber, that connect the chromatid to one of the  spindle poles.  K-fiber dynamics is 
substantially  slower than single microtubules, \cite{Inoue1995,Betterton2013}.
There are 4 principle forces: a drag force (drag coefficient $\gamma$) and an  antipoleward polar ejection force (strength $\alpha$) acting on the chromosomes,  a centromeric 
spring (spring constant $\kappa$, natural length $L$) connecting the sister kinetochores, and a pushing/pulling force $f_\pm$ from a polymerising ($+$), depolymerising ($-$)  attached K-fiber, Fig. 1C. Force balance  then give the dynamics, (intersister distance $\Delta = x_1-x_2$), \cite{Armond2015plos}
\beqa
 {d x_1 \over dt} &=& -{f_{\sigma^1(t)}\over \gamma}-{\kappa \over \gamma} (\Delta -L)- {\alpha \over \gamma} x_1 + \sqrt{2 D} \xi_1,\cr
 {d x_2 \over dt} &=& {f_{\sigma^2(t)} \over \gamma}+{\kappa \over \gamma} (\Delta -L)- {\alpha \over \gamma} x_2+  \sqrt{2 D} \xi_2, \label{eqn:dynamics}
\eea
where  $x_1, x_2$ are the  positions of the two sister 
kinetochores (normal to the spindle equatorial plane). Sisters are orientated such that  $x_1>x_2$ on average, \ie sister 1 is attached to the spindle pole to the right, sister 2 to the left. $\xi_k$ are standard white noise modelling thermal and active noise components acting on the kinetochores, parametrised by an effective diffusion coefficient $D$.
These forces have been inferred from kinetochore tracking data \cite{Burroughs2015elife,Armond2015plos}, demonstrating that the spring is in low tension throughout the oscillation compared to the dominant force generated by depolymerising microtubules.
Sister direction is labelled by $\sigma^k(t) \in \{ -,+\}$ for sister $k=1,2$.
The saw-tooth like oscillations then correspond to a sequence of \textit{coherent} sections
where the sisters move in the same direction, one attached to polymerising ($\sigma^k=+$), the other depolymerising ($\sigma^k=-$) microtubules, separated by a fast double switching event where both sisters switch direction (in either order), \ie after the first sister switches there is an \textit{incoherent} section when both sisters are either attached to polymerising (LIDS) or depolymerising (TIDS) microtubules.




The tension-clock hypothesis, \cite{Burroughs2015elife}, provides a mechanism to control directional switching. 
Analysis of (inferred) force profiles (in HeLa cells, \cite{Armond2015plos, Burroughs2015elife}, and RPE1 cells \cite{harrison2021}) provides evidence of 
 sister direction  arising from  ageing/degradation of the K-fibre from the time since the last switch. 
Specifically,  polymerising K-fibres are proposed to require increasing pulling forces to remain polymerising, while depolymerising K-fibres have a decreasing load threshold triggering rescue events,  Fig. 1C, analogous to observations of yeast kinetochore interactions with single microtubules \cite{Akiyoshi2010}.

We implement the tension-clock dependence as a (linearly)  time-varying tension threshold for each K-fibre, a sister switching direction when the spring tension hits its threshold, Fig. 1C. At a switching event we assume a K-fibre is restored to a 'prime' state with tension thresholds $T_{pol}^0$, $T_{depol}^0$, for polymerising, respectively depolymerising K-fibre, these thresholds then evolve linearly over time (age)
$
T_{pol}(t)= T_{pol}^0+a_+ (t-t^*), \ T_{depol}(t)= T_{depol}^0-a_- (t-t^*),
$
where 
$t^*$ is the time that the sister last switched  with rates  $a_+,a_-\geq 0$.
Thus, a coherent state survives provided $T_{pol}(t) < T(t) < T_{depol}(t)$, with (rescaled) spring tension  $T=\kappa (\Delta -L)/\gamma$, and ends when one of these thresholds is reached, thereby switching the associated sister, Fig. 1C. 
For incoherent states, the sister that previously switched earlier (the older sister) will reach its threshold first.
This is  a piecewise-smooth  dynamical system \cite{dibernardo,Makarenkov2015}, with reset maps on the threshold switching condition manifolds. 
Newton's cradle is a familiar exemplar of such a system. 

We initially analyse the deterministic version of Eqn. (\ref{eqn:dynamics}) with $D=0$. 
We solve Eqns. (\ref{eqn:dynamics}) 
during sections of constant polymerisation or depolymerisation. 
The intersister distance $\Delta=x_1-x_2$ decouples,
\beq
{d \Delta \over dt \,} = -{(f_{\sigma^1}+f_{\sigma^2}) \over \gamma} + 2 {\kappa L \over \gamma} - {(2 \kappa +\alpha )\Delta \over \gamma}. 
\ee
Between switching events we have the analytical solution $\Delta (t)= \Delta^*_{\sigma^1 \sigma^2} (1-e^{- (2 \kappa +\alpha) (t-t^*)/\gamma} ) + \Delta_{t^*} 
e^{- (2 \kappa +\alpha) (t-t^*)/\gamma}$, where $\Delta_{t^*}$ is the separation at previous switching  time  $t^*$. Here $\Delta^*_{\sigma^1 \sigma^2}=(2\kappa L-(f_{\sigma^1}+f_{\sigma^2}) )\big/(2 \kappa + \alpha),$
corresponding  to the steady states for the fixed  directional  states $\sigma=\{++,+-,--\}$.
Since the spring force is a function of $\Delta$ only,
switching dynamics is  only a function of the $\Delta$ dynamics.

Periodic solutions can be constructed
by suitably combining sections of these solutions through a sequence of 
states $\sigma$, giving oscillatory solutions in the $(\Delta,T^1, T^2)$ space, where $T^k(t)$ is the threshold for sister $k$.
Oscillatory solutions are defined by their switching choregraphy, and the times $t_i$, $i=1..n$, of the $n$ sections.
There are at least 5 periodic solutions (subscript denoting number of sections), Fig. 2A-C: 

\begin{enumerate}

\item {\bf LIDS$_2$ (TIDS$_2$) oscillation}: an oscillation with  LIDS (TIDS) choreography throughout.

\item {\bf BKT$_4$ oscillation}:
Biased kinetochore oscillation where one of the sisters always switches first (alternate LIDS, TIDS events).

\item {\bf Asymmetric  LIDS$_4$ oscillation}:  LIDS choreography throughout with asymmetric sections. 

\item
{\bf Breather}, with  anti-correlated sisters.
\end{enumerate}

We performed a stability analysis and bifurcation analysis of these solutions
in the $a_+ , a_-$ phase plane. 
These oscillatory solutions occur only in particular regions of the $a_+ , a_-$
phase plane, Fig. 2D. TIDS and BKT oscillations are always stable,  LIDS$_4$ unstable and  LIDS$_2$ can be either stable or unstable.
The system
exhibits standard 
bifurcations and border collision bifurcations, additional bifurcations of smooth piecewise dynamical systems \cite{dibernardo,Makarenkov2015}, where the fixed point collides with the switching manifold. Specifically we have the following collision surfaces:

\begin{itemize}

\item $\Sigma_{LIDS}$. 
  On this surface LIDS$_2$ and LIDS$_4$ solutions collide, i.e. there are continuations in
  parameter space for both solutions. This is a standard bifurcation;
there is a   
  change in stability of LIDS$_2$  in a period  doubling bifurcation (the period of $\Delta$ doubles between LIDS$_2$ and LIDS$_4$). 

  \item
$\Sigma_{\Delta_{+-}}$. 
    On this surface LIDS$_2$, TIDS$_2$  and KBT solutions meet when an incoherent section time $\rightarrow 0$ (a joint switch) and the kinetochores are stationary,  $x_{1,2}=\pm \Delta_{+-}/2$ whilst only $T_k(t)$ oscillate.
    This is a border collision bifurcation  where a fixed point collides with the switching manifold \cite{Makarenkov2015}; on this surface the threshold conditions change, the sister who switches swops over, eg LIDS$_2$ $\sigma=(+-,++,-+,++)$ to TIDS$_2$ $\sigma=(+-,--,-+,--)$. There is no change in
    stability.  

    
\item 
$\Sigma_{LIDS4:BKT}$. 
  On these surfaces a BKT (either sister)
  and a LIDS$_4$ solution meet as an incoherent section time $\rightarrow 0$. 
  This is a border collision bifurcation. 
  

\end{itemize}

A (schematic) bifurcation diagram in $a_-$, Fig. 2E,  summarises the relationaship of these solutions and their oscillation format. The
period of the oscillations clearly demonstrates the multiple oscillations that exist in various parts of this bifurcation diagram, Fig. 2F.


The breather solution  occurs everywhere but is unstable. This is because joint switching is unstable - consider  a small perturbation of the breather oscillation that separates the thresholds. When one of the sisters switches its direction, the spring tension  moves  in the opposite direction, away from the threshold of the second sister.

For human cells the oscillation period is around 70s, and the amplitude 
of the order of a micron, \cite{Armond2015plos}. The model is able to reproduce such values
with $a_+, a_-$ of the order of $10^{-4}$ $\mu$m/s$^2$ (with threshold initialisation $T^0_{pol/depol}$ as Fig. 2).
The intersister distance $\Delta$ oscillates with  double the sister oscillation frequency, \cite{Jaqaman2010},  
(as determined by autocorrelation), thus highly asymmetric oscillations BKT$_4$ are not observed. 
However, experimentally oscillations that comprise only single types of switching events are rare, and a typical oscillatory trajectory
has both LIDS and TIDS events (the LIDS/TIDS ratio depends on cell line, 1.56 \cite{Burroughs2015elife}, 3.76  \cite{Armond2015plos}, 2.22 \cite{harrison2021}). This indicates system noise is needed as the
 deterministic oscillations with one type of oscillation choreography (with 0, 50 or 100\% LIDS events depending on region) 
 are too uniform in structure.


With diffusive noise, the intersister distance $\Delta$ satisfies the Ornstein-Uhlenbeck equation, and the switching time is given by solving
a first passage time (FTP) problem with time varying boundaries. No analytical solution for the FPT problem is 
known in this case, and a numerical method is used \cite{yi2010}.
In the presence of diffusive noise we find that the 
 oscillation regions in the phase diagram deteriorate with increasing noise,
Fig. 3A-D. 
  At low noise, $D=10^{-6}$ $\mu$m$^2$/s, the LIDS and KBT regions are essentially intact, but the TIDS region deteriorates with KBT switching (50:50 LIDS/TIDS events) occurring in the proximity of the bifurcation surface. As noise
increases  the 
breather oscillation becomes metastable, 
the breather  oscillation being the dominant state in most regions of the phase diagram at  high noise $D=2 \times 10^{-4}$ $\mu$m$^2$/s, Fig. 3C,D, where
only the LIDS oscillation  remains intact at low $a_+$.
Trajectories acquire transient periods of breather oscillation at high noise, $D=2 \times 10^{-4}$ $\mu$m$^2$/s, Fig. 3G.
Noise induced stabilisation of unstable structures occurs in  other systems \cite{samoilov2005,Turcotte2008,Rue2011,wu2022}. 

This deterioration of saw-tooth like oscillations by noise induced stabilisation of the breather state can be suppressed 
by preventing  switching events occurring too frequently, \ie if there
is a reset time $t_{ref}$ after a switching event when further switching of that sister is prohibited. 
Since the breather state has a short period, this suppresses this state.
With a 15s reset period we observe oscillations with a mix of LIDS and TIDS switches and the loss of transient breather states, Fig 3H.

We have presented a new mathematical model  
of metaphase chromosome oscillations in mammals, where microtubule (bundle) load dependent directional switching is controlled by an ageing threshold, a control
mechanism we refer to as the tension-clock.
The phase diagram of the deterministic system is derivable semi-analytically.
The system displays  rich dynamics with at least 5 oscillatory solutions, whilst  the unstable breather state is stabilised on introduction of noise. 
This mechanism reproduces realistic oscillations,  in particular reproducing 
mixed LIDS/TIDS choreographies with realistic amplitudes and
periods, Fig. 3H, and has low intersister tension. 
The next step is to infer the tension-clock parameters from experimental data. Given the semi-Markov nature of  the noisy tension-clock model
this is challenging.

\vskip 1cm

\bibliography{tenclock_refs}

\vskip 1cm
\noindent
\textbf{Figure Legends}

{\small Figure 1. The tension-clock model of chromosome oscillations.
\textbf{A}. Typical oscillatory dynamics of a pair of human kinetochores annotated for LIDS and TIDS, see  \cite{Burroughs2015elife} for methods.
\textbf{B}. 
Inferred spring force through a LIDS (magenta, 1614 events) or TIDS (black, 449 events) event, showing relaxation, respectively overstretch of the spring force after the switching event.
\textbf{C}. Schematic of the tension-clock mechanical model.
Chromosome movement is dependent on forces generated by microtubules, \cite{mcintosh2012,nenad2016}. 
Here a pair of sister kinetochores have
microtubule attachments to respective spindle poles. Microtubules, bundled in K-fibres,
are assumed to have an evolving threshold, and switch direction if that threshold is met, upper panel.
{\bf A, B} reproduced from \cite{Armond2015plos} with permission.

Figure 2. Dynamic behaviour of the deterministic tension-clock model.
Simulations of the tension-clock model showing
\textbf{A.} Breather state (coincident switching of sisters), 
\textbf{B.} LIDS oscillation, \textbf{C.} Biased KT oscillation (BKT, sister 1 always switching). Directional switching events are shown, first sister switching marked
with  cyan for LIDS, magenta for TIDS.
\textbf{D.} Trajectories in $x_1-x_2$ plane.
\textbf{D.} Phase plot showing existence of LIDS, TIDS and BKT oscillations in the $a_+, a_-$ phase plane.
\textbf{E.} Bifurcation diagram for $a_-$ constant cross section of D, with schematic of trajectory phase plane (in $x_1-x_2$).
Bifurcations: pitchfork bifurcations, red, purple, border collision bifurcation blue. 
The solutions continue (dashed) beyond the border surface as unphysical solutions, \ie there is  a
    solution with the same switching choreography but the second sister switches upon the second meeting of the threshold condition (from the wrong direction).
    \textbf{F} Period of oscillatory solutions with $a_-$ ($a_+=5 \times 10^{-4}$; stable oscillations shown with solid lines).
       Mechanical parameters are set to $f_+/\gamma=15$ nm/s, $f_-/\gamma=-5$ nm/s, $\kappa/\gamma=0.025$ s$^{-1}$, $\alpha/\gamma=0.01$ s$^{-1}$, typical for oscillatory trajectories, \cite{Armond2015plos}. Thresholds are initialised
         to $T^0_{pol}=-0.1$ nm/s, $T^0_{depol}=12$ nm/s. 
       $a_\pm$ in $\mu$m/s$^2$ in {\bf D, E, F}.

Figure 3. The stochastic tension-clock model.
{\bf A/C}. Fraction of time that  sisters are coherent as a function of the threshold gradients  $a_\pm$
for {\textbf A}. low noise $D=10^{-6}$ $\mu$m$^2$/s, {\textbf C}. high noise $D=1.25 \times 10^{-4}$ $\mu$m$^2$/s. 
\textbf{B/D}. Fraction of switching events that are LIDS for cases {A/C}.
{\bf E/F/G}. Typical trajectories in the BKT oscillation region of Fig 2D ($a_+= 3 \times 10^{-4}, a_-=  2 \times 10^{-5}$) 
with increasing diffusive noise: {\textbf E}. low
diffusive noise  $D=10^{-6}$, {\textbf F}. medium diffusive noise  $D=10^{-5}$, {\textbf G}. high diffusive noise  $D= 10^{-4}$.
{\bf H} Stochastic tension-clock model with a reset period of 15 s and other parameters as {G}. 
In \textbf{E-H} LIDS events are shown in cyan, TIDS in magenta.
Other parameters as Fig. 2.
$a_\pm$ in $\mu$m/s$^2$.

\vskip 1cm

\vfil
\eject
\begin{figure}
\includegraphics{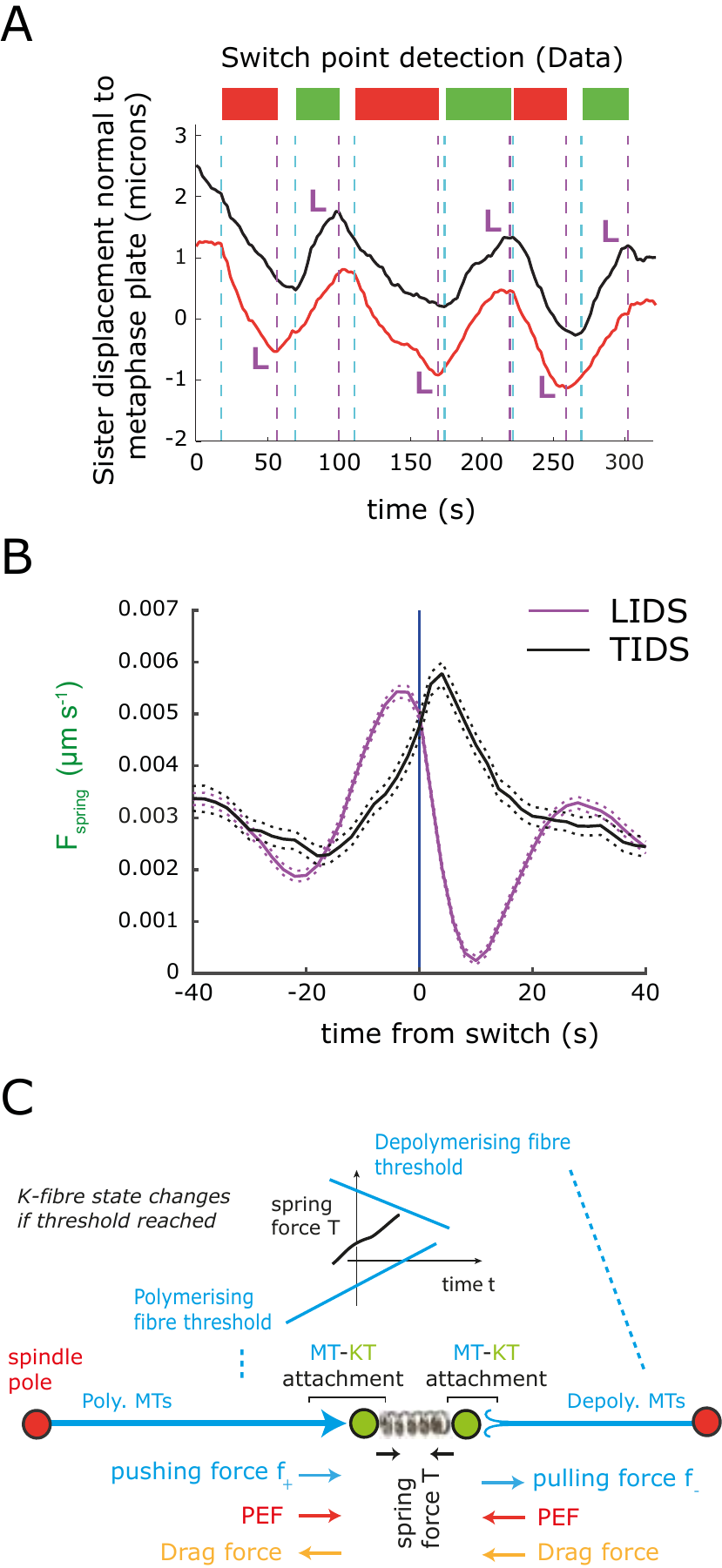}
\caption{\ }
\end{figure}

\eject
\begin{figure*}
\includegraphics{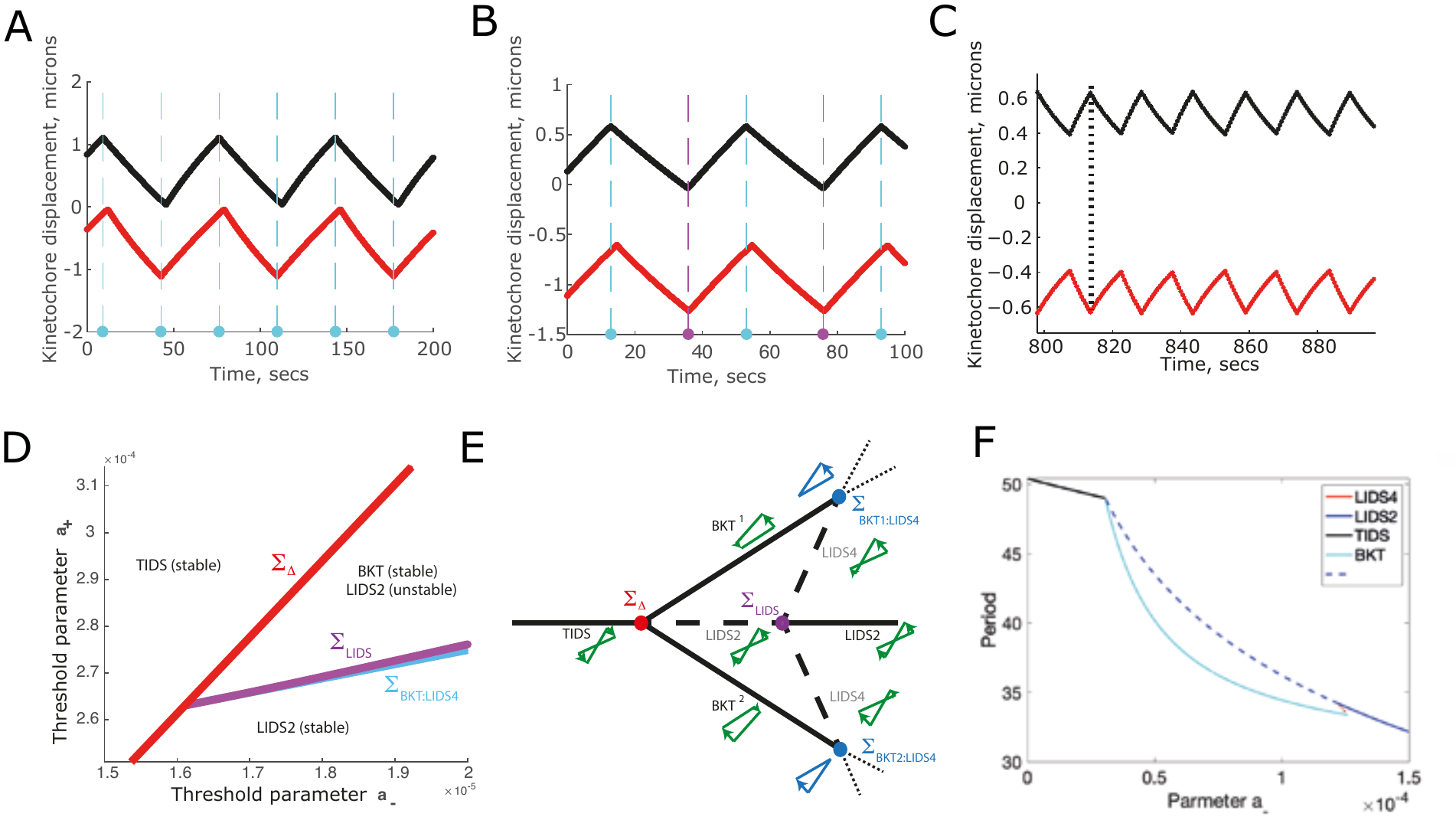}
\caption{\ }
\end{figure*}

\eject
\begin{figure*}
\includegraphics{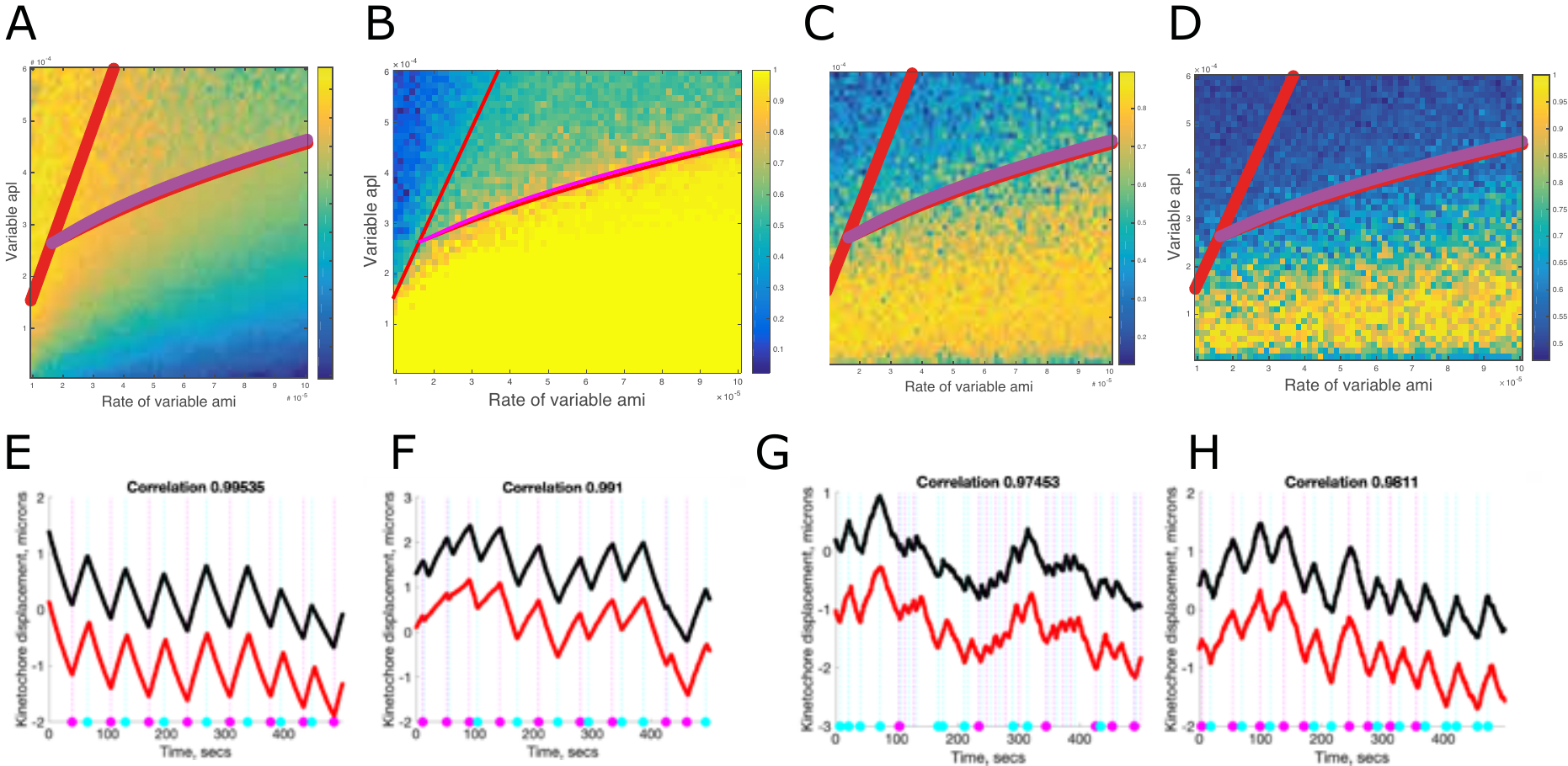}
\caption{ \ }
\end{figure*}

\end{document}